\documentclass[a4paper,11pt]{article}
\usepackage{jheppub}
\usepackage{subfig}
\usepackage{amsmath}
\usepackage{amssymb}
\usepackage{latexsym}

\newcommand{\VV}{ }

\newcommand{\sss}[1]{\scriptscriptstyle#1}

\newcommand{\as}{a_s}
\newcommand{\al}{\alpha}
\newcommand{\be}{\beta}

\newcommand{\Tr}{{\mathrm{ Tr}}}
\newcommand{\prd}{\partial}

\newcommand{\ed}{\end{document}}

\newcommand{\MSbar}{\ensuremath{\overline{\text{MS}}}}

\newcommand{\g}{\gamma}

\newcommand{\ovl}[1]{\overline{#1}}
\newcommand{\ice}[1]{\relax}
\newcommand{\re}[1]{(\ref{#1})}

\newcommand{\beq}{\begin{equation}}
\newcommand{\eeq}{\end{equation}}
\newcommand{\bea}{\begin{eqnarray}}
\newcommand{\eea}{\end{eqnarray}}

\newcommand{\ba}{\begin{array}}
\newcommand{\ea}{\end{array}}

\newcommand{\dmu}{\mu^2\frac{d}{d\mu^2}}

\newcommand{\sbz}{  }

\let\*=\,

\newcommand{\dd}{{\rm d}}

\newcommand{\nnb}{\nonumber}

\newcommand{\Tf}{T }

\newcommand{\trAnlA}{\Tf\, n_f }

\newcommand{\cfAcaA}{C_F\, C_A}
\newcommand{\cfAtrAnlA}{C_F\, T\, n_f}
\newcommand{\cfB}{C_F^2}

\newcommand{\cfC}{C_F^3 }
\newcommand{\cfAtrBnlB}{\Tf^2\, n_f^2\,C_F}
\newcommand{\cfBtrAnlA}{\Tf\, n_f\,C_F^2}
\newcommand{\cfAtrAnlAcaA}{\Tf\, n_f\,C_F\, C_A}
\newcommand{\cfBcaA}{C_F^2 \,C_A}
\newcommand{\cfAcaB}{C_F \,C_A^2}

\newcommand{\cfD}{C_F^4 }
\newcommand{\dFFdRinvAnlA}{n_f\,\frac{d_F^{abcd}\,d_F^{abcd}}{d_R}}
\newcommand{\dFARinvA}{\frac{d_F^{abcd}\,d_A^{abcd}}{d_R}}
\newcommand{\cfAtrCnlC}{\Tf^3\, n_f^3\,C_F }
\newcommand{\cfBtrBnlB}{\Tf^2\, n_f^2\,C_F^2}
\newcommand{\cfAtrBnlBcaA}{\Tf^2\, n_f^2\,C_F \,C_A}
\newcommand{\cfCtrAnlA}{\Tf\, n_f\,C_F^3}
\newcommand{\cfBtrAnlAcaA}{\Tf\, n_f\,C_F^2 \,C_A}
\newcommand{\cfAtrAnlAcaB}{\Tf\, n_f\,C_F \,C_A^2}
\newcommand{\cfCcaA}{C_F^3 \,C_A}
\newcommand{\cfBcaB}{C_F^2 \,C_A^2}
\newcommand{\cfAcaC}{C_F \,C_A^3}

\def\beq{\begin{equation}}
\def\eeq{\end{equation}}
\def\bea{\begin{eqnarray}}
\def\eea{\end{eqnarray}}
\def\bq{\begin{quote}}
\def\eq{\end{quote}}

\def\nnb{\nonumber}

\def\nnb{\nonumber}

\def\ba{\begin{array}}
\def\ea{\end{array}}

\title{
\boldmath 
Vector Correlator in Massless QCD at  Order ${\cal O}(\alpha_s^4)$ and the QED $\beta$-function at five loop}
\author[a]{P.~A.~Baikov,}
\author[b]{K. G. Chetyrkin,}
\author[b]{J.~H.~K\"uhn,}
\author[b]{and J.~Rittinger}

\affiliation[a]{
Skobeltsyn Institute of Nuclear Physics, Lomonosov Moscow State University, 
1(2), Leninskie gory, Moscow 119234, Russian Federation
        }
\affiliation[b]{Institut f\"ur Theoretische Teilchenphysik, Karlsruhe
  Institute of Technology (KIT), Wolfgang-Gaede-Stra\ss{}e 1, 726128 Karlsruhe, Germany}

\emailAdd{baikov@theory.sinp.msu.ru}
\emailAdd{Konstantin.Chetyrkin@kit.edu}
\emailAdd{johann.kuehn@kit.edu}
\emailAdd{rittinger@particle.uni-karlsruhe.de}

\abstract{
We present a  concise summary of recent results for the
vector correlator in  massless QCD at order ${\cal O}(\alpha_s^4)$,
with all colour factors being given for a generic colour group. As a
direct consequence we arrive at: (i) the full QCD contribution to the
QED $\beta$--function of order $\alpha^2\, \alpha_s^4$ in the \MSbar- \  and
MOM-schemes; (ii) the full five-loop result of order $\alpha^6$ for
the $\beta$-function of QED with a generic number of single-charged
fermions, again  for  the \MSbar- \  and MOM-schemes.
}


\begin{document}

\hfill SFB/CPP-12-36, TTP12-18

\maketitle

\section{Introduction \label{sec:intro}}

Quantum electrodynamics can be considered to be one of the most
successful theoretical concepts, allowing to predict experimental
results with unchallenged precision. Here the anomalous magnetic moment
of the electron (see, \cite{Aoyama:2012wj} and references therein),
positronium spectroscopy  \cite{Karshenboim:2005iy} or 
properties of light hydrogenic bound states \cite{Eides:2007}
may be listed as most outstanding examples. At the same time the detailed investigation of its
mathematical structure has lead to fundamental field-theoretical
concepts, like renormalization or the renormalization group with
anomalous dimensions and the $\beta$-function as important elements for
the analysis of its high energy behaviour \cite{Stueckelberg53,GellMann:1954fq,Bogolyubov:1956gh}.

As a consequence of Ward identities the $\beta$-function of QED is closely
related to the vacuum polarization function $\Pi(q^2)$ and, in contrast to
non-abelian gauge theories like QCD, no vertex corrections need to be
evaluated. Indeed, the coefficient of the $n$-th term of the perturbative
series of the $\beta$-function can be directly obtained from the
absorptive part of $\Pi(q^2)$ at the same order.
The same absorptive part,
evaluated including QCD corrections, can be employed to predict the
Adler function \cite{Adler:1974gd} and thus the total cross section for electron
positron annihilation into hadrons or the $\tau$-lepton semileptonic
decay rate. The precise experimental information for these fundamental
quantities has stimulated dedicated projects for the evaluation of 
three- \cite{Chetyrkin:1979bj}, four- \cite{Gorishnii:1990vf,Surguladze:1990tg} 
and recently even five-loop corrections  
\cite{Baikov:2001aa,Baikov:2008jh,Baikov:2010je,Baikov:2012er,Baikov:dummy}
for these fundamental quantities. Conversely,  it is obvious that the information
contained in these five-loop QCD results is sufficient to evaluate the
QED vacuum polarization and the $\beta$ function in the same five-loop
order. Partial results in this direction have been presented in \cite{Baikov:2001aa,Baikov:2008cp,Baikov:2010je}.

It is the purpose of this work  to collect all available  results for the
vector correlator in  massless QCD in a form  suitable 
for a complete evaluation of the QED $\beta$ function. Two cases will be
considered: the QCD contribution to the QED $\beta$-function of order $\alpha^2\alpha_s^4$ and the QED $\beta$-function of
order $\alpha^6$ for a generic number of charged fermions.

For readers's convenience  we provide in the paper the   complete results for the vector 
current correlator, related  $\beta$-functions and  anomalous dimensions, including  also
lower order contributions. As the  latter are known since long,  we refer the reader
to works \cite{Chetyrkin:1996ia,Gorishnii:1990kd,Erler:1998sy} for corresponding historical discussions
and citations.  Our full results are also available  (in computer-readable form) in
{\tt http://www-ttp.physik.uni-karlsruhe.de/Progdata\\/ttp12/ttp12-018.}

\section{Preliminaries \label{sec:pre}}
%
Our main object will be the polarization  function $\Pi(L,\as)$ of the flavor singlet vector current defined 
as
\beq
 (-g_{\al\be} q^2 + q_{\al} q_{\be}) \, \Pi(L,\as) =
i\int\dd^4 xe^{iq\cdot x}\langle 0|{\rm  T}j_\al(x)j_\be(0)|0\rangle
{},
\label{Pi}
\eeq
with $j_{\al} = \sum_i\ovl{\psi}_i\gamma_{\al}\psi_i$ and  $Q^2 = -q^2$,
$L = \ln \frac{\mu^2}{Q^2}$,  $\as = \frac{\alpha_s(\mu)}{\pi}$.
We work within massless QCD with $n_f$ quark flavours,  $\mu$ is the normalization point of the \MSbar-scheme.

\begin{figure}[h]
\begin{center}
\subfloat[]{ \includegraphics[height=2.3cm]{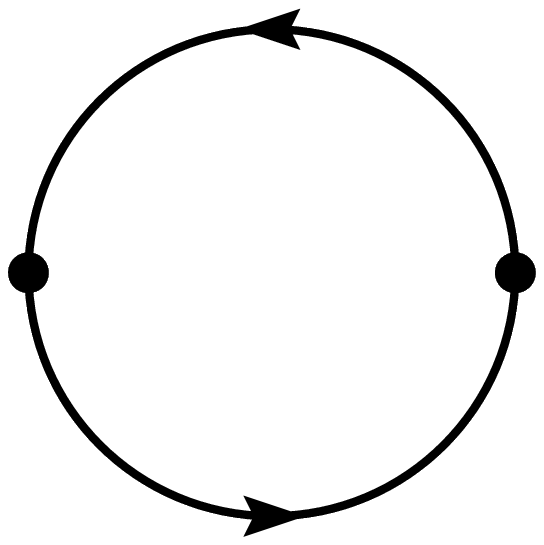} }
\subfloat[]{ \includegraphics[height=2.3cm]{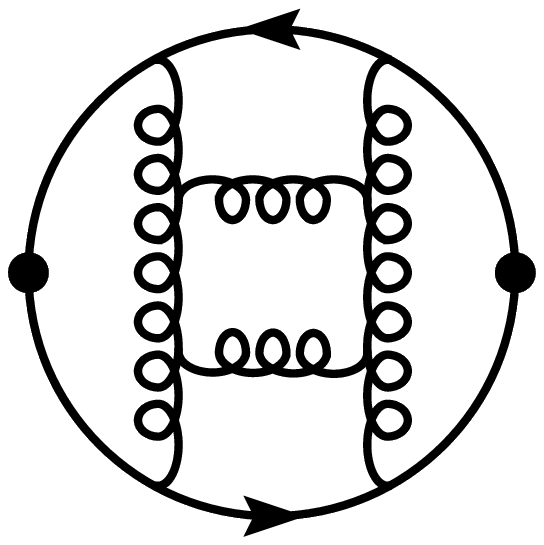} }
\subfloat[]{ \includegraphics[height=2.3cm]{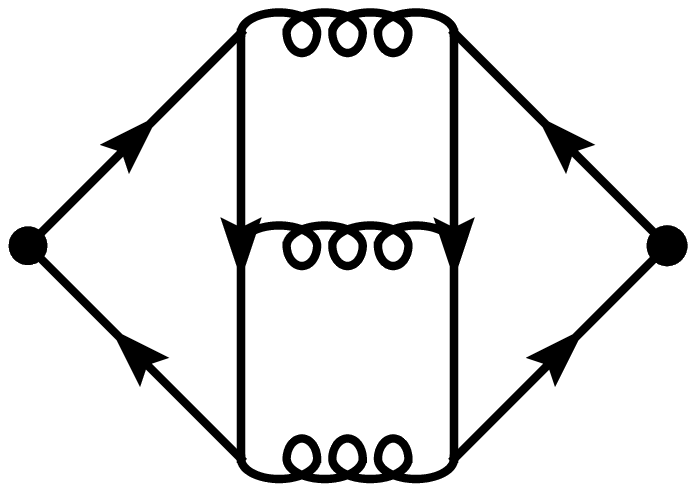} }
\subfloat[]{ \includegraphics[height=2.3cm]{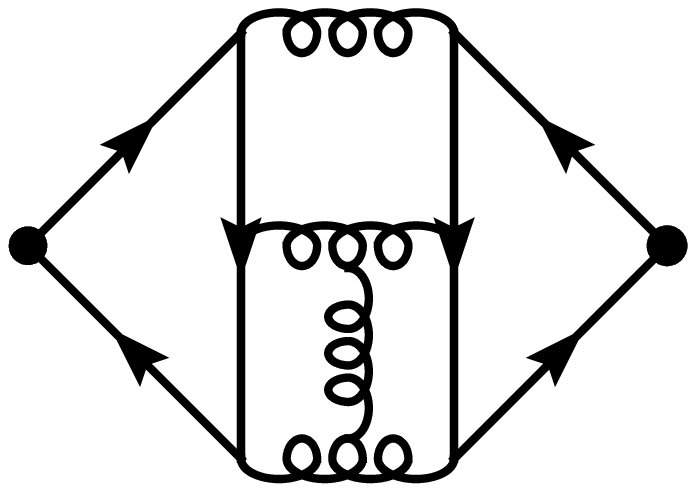} }
\end{center}
\caption{Examples of a lowest order and five-loop non-singlet  (a),(b) and singlet (c),(d)  diagrams
contributing to the vector correlator.
}
\end{figure}

All diagrams contributing to  $\Pi(L,\as)$ can be  naturally decomposed into two classes: singlet and
non-singlet ones. The non-singlet diagrams are those with  both external currents belonging to
one and the same quark loop; all other diagrams are referred to as singlet ones (see Fig. 1). 
As a consequence of  this classification  the full  polarization  function can be presented as follows:
\beq
\Pi(L,\as) = n_f\, \Pi^{\sss NS}(L,\as) +  n_f^2\,\Pi^{\sss SI}(L,\as)
{}.
\eeq
It is worth to note  that the knowledge of $\Pi^{\sss NS}$ and $\Pi^{\sss SI}$ is  enough  to construct
the polarization function arising from  two vector currents with generic flavour structure. Indeed, if
$j_\al^a =  \ovl{\psi}\gamma_{\al}t^a \psi$ and $j_\al^b =  \ovl{\psi}\gamma_{\al}t^b \psi$
($t^a$ and $t^b$ being two arbitrary matrixes in  flavor space) then the corresponding
polarization function $\Pi^{a b}$ is, obviously,  given by the relation
\[
\Pi^{a b} = \Tr(t^a\, t^b)\,\Pi^{\sss NS} +  \Tr(t^a)\,\Tr(t^b)\, \Pi^{\sss SI}
{}.
\]
For instance, the polarization function corresponding to the   electromagnetic vector current
\[
j^{\sss EM}_{\mu} = \sum_i q_i \ovl{\psi}_i\gamma_{\mu}\psi_i
\]
($q_i$ stands for the electric charge of
the quark field $\psi_i$)  is given by the expression:
\beq
{\Pi}^{\sss EM} = \left(  \sum_i q_i^2 \right) \Pi^{\sss NS} + 
\left(  \sum_i q_i \right)^2
 \, \Pi^{\sss SI}
\label{PiEM}
 {}.
\eeq
The absorptive part of  ${\Pi}^{\sss EM}$ is  directly related to the important physical quantity
$R(s)\equiv \sigma(e^+e^-\to {\rm hadrons}) / \sigma(e^+e^-\to\mu^+\mu^-)$ via the relation
\beq
R(s) = 6 \pi\, \left( \Pi^{\sss EM}(-s - i\epsilon) - \Pi^{\sss EM}(-s + i\epsilon) \right)
\label{R}
{}.
\eeq
Another useful  object -- so-called Adler function --- is defined as 
\cite{Adler:1974gd} 
\beq
{D}(L,a_s) =  -12\, \pi^2
Q^2\, \frac{\mathrm{d}}{\mathrm{d} Q^2} \Pi(L=\ln\frac{\mu^2}{Q^2},a_s)
{}.
\label{D}
\eeq
Note that the choice of the  coefficients $6 \pi$ and 
$-12\pi^2 $ in eqs. \re{R} and \re{D} is conventionally fixed by the  requirement that
in  Born approximation 
\[
R(s) = \left(  \sum_i q_i^2 \right) d_R\,\left(1+ {\cal O}(\alpha_s)\right), \ \ 
D(L,a_s) = n_f \,d_R \,\left(1+ {\cal O}(\alpha_s) \right)
{},
\]
where the $d_R$ is the dimension of the quark representation of the colour gauge group
($d_R =3 $ for QCD).  

In spite of the fact, that the vector current in QCD is a
scale-invariant object (that is it has zero anomalous dimension),  the
polarization function is {\em not}  due to the short distance singularities
of the T-product in eq.~\re{Pi}. The corresponding evolution equation
reads (see, e.g. \cite{Chetyrkin:1996ia})
\beq
\label{rgea}
\left(
\mu^2\frac{\partial}{\partial\mu^2}
 + 
\beta(a_s) 
\frac{\partial}{\partial a_s} 
\right)
           \Pi(L,\as) =   \gamma^{\rm \VV }(a_s)
{}
\eeq
or, equivalently,  
\beq
\frac{\prd }{\prd L}\,  \Pi (L,\as) =
\g^{\rm \VV }(a_s)
-
 \beta( a_s) \frac{\prd }{\prd a_s}
\Pi(L,\as)
\label{rgPi2}
{}.
\eeq
Here 
the anomalous dimension  $\gamma^{\rm \VV }$
is a series in $\as$ of the form
\beq
\gamma^{\rm \VV }(a_s) = \frac{1}{16\,\pi^2}\sum_{i \ge 0} \gamma_i^{\VV } \as^i
\eeq
and the QCD $\beta$-function is defined as
\beq
\dmu a_s  =  \beta(a_s) \equiv
-\sum_{i\geq0}\beta_i a_s^{i+2} 
\label{beta:def}
{}.
\eeq
For our analysis  we need to know the QCD $\beta$-function with three-loop accuracy, the corresponding
result is known since long from \cite{Tarasov:1980au,Larin:1993tp}.   
Note that the ratio $R(s)$  and the  Adler function are scale-invariant as the rhs of eq. \re{rgea}, being 
an anomalous dimension does not depend on either  $Q$, $\mu$ or $L$.

In massless QCD   eq.~\re{rgPi2} directly leads
to an important  relation for the Adler function:
\beq
D(L,a_s) =  12\pi^2\, \left(\g^{\rm \VV }(a_s)
-\left(
 \beta( a_s) \frac{\prd }{\prd a_s}
\right) \Pi(L,\as)
\right)
\label{Adler:master:eq}
{}.
\eeq
{E}qs.~\re{rgPi2} and \re{Adler:master:eq}, are crucial  for our ability
to compute the Adler function and $R(s)$ at order $\alpha_s^4$ which, in principle, are 
 determined  by {\em five-loop } diagrams. The simplification arises  because
the rhs of eq.~\re{Adler:master:eq} is, in fact, expressible {\em
exclusively} through {\em four-loop} massless propagators.  First,
only the four-loop ${\cal O}(\alpha_s^3)$ approximation to $\Pi(L,a_s)$ is
required as the  $\beta$-function starts from order
$\alpha_s^2$. Second, the evaluation of any $(L+1)$-loop anomalous
dimension in the \MSbar-scheme  can be reduced, with the help of the $R^*$-operation \cite{Chetyrkin:1984xa}, to the
evaluation of some $L$-loop masslesss propagators \cite{Chetyrkin:1996ez}. 
Finally,  four-loop massless propagators can be  reduced to 28 master integrals. The reduction
is based on evaluating  
sufficiently many terms of the $1/D$ expansion \cite{Baikov:2005nv} of
the corresponding coefficient functions \cite{Baikov:1996rk}. The master integrals are known analytically from
\cite{Baikov:2010hf,Lee:2011jt}.

Needless to say  that any direct way of computing 5-loop diagrams, contributing,  for instance, to the 
Adler function {\em without} the  use of \re{Adler:master:eq} is hopelessly  difficult and, presumably, will stay so 
for years and years   ahead.

\section{Results for the vector correlator}

In this section we present all currently available results for the
polarization function $\Pi$  and the anomalous dimension $\g^{\VV }$ which have
been utilized in works \cite{Baikov:2008jh,Baikov:2010je,Baikov:2012er,Baikov:dummy}
to construct the Adler function and the
ratio $R(s)$ in massless QCD to order $\alpha_s^4$.

\subsection{The polarization function $\Pi$}
By presenting  the perturbative expansion of the  polarization function $\Pi$ as follows
\beq
\Pi^{\sss NS} =  \frac{d_R}{16\pi^2 }\,\left( \sum_{i \ge 0}
p^{\sss NS}_i \, a_s^i
\right),
\ \ 
\Pi^{\sss SI} =  \frac{d_R}{16\pi^2 }\, \left( \sum_{i \ge 3}
p^{\sss SI}_i \, a_s^i
\right)
{},
\eeq 
we get
\bea
p_0^{\sss NS} &=&   \frac{20}{9}
\label{p0ns}
{},
\\
p_1^{\sss NS} &=&   
C_F
\left[
\frac{55}{12} 
-4  \sbz \zeta_{3}
\right]
{},
\label{p1ns}
\\
p_2^{\sss NS} &=&   
{}\cfB
\left[
-\frac{143}{72} 
-\frac{37}{6}  \sbz \zeta_{3}
+10  \sbz \zeta_{5}
\right]
{+}\cfAcaA
\left[
\frac{44215}{2592} 
-\frac{227}{18}  \sbz \zeta_{3}
-\frac{5}{3}  \sbz \zeta_{5}
\right]
\nnb
\\
&{}&\hspace{0.7cm} {+}\, \cfAtrAnlA
\left[
-\frac{3701}{648} 
+\frac{38}{9}  \sbz \zeta_{3}
\right]
{},
\label{p2ns}
\eea
\bea
p_3^{\sss NS} &=&   
{}\cfC
\left[
-\frac{31}{192} 
+\frac{13}{8}  \sbz \zeta_{3}
+\frac{245}{8}  \sbz \zeta_{5}
-35  \sbz \zeta_{7}
\right]
\nonumber
{+} \cfAtrBnlB
\left[
\frac{196513}{23328} 
-\frac{809}{162}  \sbz \zeta_{3}
-\frac{20}{9}  \sbz \zeta_{5}
\right]
\nonumber\\
&{}& \hspace{0.7cm}{+}\cfBtrAnlA
\left[
-\frac{7505}{10368} 
+\frac{1553}{54}  \sbz \zeta_{3}
-4  \,\zeta_3^2
+\frac{11}{24}  \sbz \zeta_{4}
-\frac{250}{9}  \sbz \zeta_{5}
\right]
\nonumber\\
&{}&  \hspace{0.7cm}{+} \cfAtrAnlAcaA
\left[
-\frac{5559937}{93312} 
+\frac{41575}{1296}  \sbz \zeta_{3}
+\frac{2}{3}  \,\zeta_3^2
-\frac{11}{24}  \sbz \zeta_{4}
+\frac{515}{27}  \sbz \zeta_{5}
\right]
\nonumber\\
&{}&\hspace{0.7cm}{+} \cfBcaA
\left[
-\frac{382033}{20736} 
-\frac{46219}{864}  \sbz \zeta_{3}
-\frac{11}{48}  \sbz \zeta_{4}
+\frac{9305}{144}  \sbz \zeta_{5}
+\frac{35}{2}  \sbz \zeta_{7}
\right]
\nonumber\\
&{}&\hspace{0.7cm}{+} \cfAcaB
\left[
\frac{34499767}{373248} 
-\frac{147473}{2592}  \sbz \zeta_{3}
+\frac{55}{6}  \,\zeta_3^2
+\frac{11}{48}  \sbz \zeta_{4}
-\frac{28295}{864}  \sbz \zeta_{5}
-\frac{35}{12}  \sbz \zeta_{7}
\right]
{},
\label{p3ns}
\end{eqnarray}
\beq
p_3^{\sss SI} = \frac{d^{abc}\,d^{abc}}{d_R}   
\left\{
\frac{431}{1728} 
-\frac{21}{64}  \sbz \zeta_{3}
-\frac{1}{6}  \,\zeta_3^2
-\frac{1}{16}  \sbz \zeta_{4}
+\frac{5}{16}  \sbz \zeta_{5}
\right\}
{}.
\label{p3si}
\eeq
Here  $C_F$ and $ C_A$ are the quadratic Casimir
operators of the fundamental and the adjoint representation of the colour Lie algebra,
$d^{abc} = 2\, \mathrm{Tr}(\{\frac{\lambda^a}{2},\frac{\lambda^b}{2}\}  \frac{\lambda^c}{2})$,
$T$ is the trace normalization of the fundamental representation. 
The exact definitions of the colour structures 
${d_F^{a b c d} d_A^{a b c d}}$ and ${d_F^{a b c d}d_F^{a b c d}}$
appearing below 
are given in \cite{Vermaseren:1997fq}.
For  QCD (colour gauge group SU(3)):
\bea
 d_R &=&3\,,\, C_F =4/3\,,\, C_A=3\,,\, T=1/2\,,\,
\nnb
\\
{d_F^{a b c d} d_A^{a b c d}} &=& \frac{15}{2}\,,\, {d_F^{a b c d}d_F^{a b c d}} = \frac{5}{12},  \ \ 
d^{abc}\,d^{abc} = \frac{40}{3}
{}.
\label{SU3}
\eea
For the  particular case of the $U(1)$ gauge group the colour factors assume the following values: 
\beq
 d_R =1\,,\, C_F =1\,,\, C_A=0\,,\, T=1 \,,\,
{d_F^{a b c d} d_A^{a b c d}} = 0\,,\, {d_F^{a b c d}d_F^{a b c d}} = 1,  \ \ 
d^{abc}\,d^{abc} = 16
{}.
\label{U1}
\eeq

Note that in eqs.~(\ref{p0ns}--\ref{p3si}) we have set to zero $L =\ln
\bigl(\mu^2/Q^2 \bigr)$. The full dependence on $L$ can be easily
restored from evolution eq.~\re{rgPi2} and the anomalous dimension
$\g$ given below.

\subsection{The anomalous dimension $\g^{\VV }$}

On decomposing the anomalous dimension $\g^{\VV }$ into  non-singlet and singlet terms
\beq
\g^{\sss \VV } = {n_f\,} \g^{\sss \VV NS} + {n_f^2}\, \g^{\sss \VV SI}, \ \ 
\g^{\sss \VV NS}  =  \frac{d_R}{16\pi^2 }\,\left( \sum_{i \ge 0}\g^{\sss \VV NS}_i\, a_s^i\right), \ \ 
\g^{\sss \VV SI}  =  \frac{d_R}{16\pi^2 }\,\left( \sum_{i \ge 3}\g^{\sss \VV SI}_i\, a_s^i\right), \ \ 
\eeq
we get
\begin{eqnarray}
\g_0^{\sss NS} &=&  \frac{4}{3}
\label{g0ns}
{},
\\
\g_1^{\sss NS} 
&{=}& C_F
{},
\label{g1ns}
\\
\g_2^{\sss NS} &=&  
\cfB
\left[
-\frac{1}{8}\right]
{+}\cfAcaA
\left[
 \frac{133}{144}\right]
{+}\cfAtrAnlA
\left[
-\frac{11}{36}\right]
{},
\label{g2ns}
\eea
\bea
\g_3^{\sss NS}  &=&  
\cfC
\left[
-\frac{23}{32}\right]
{+}\cfAtrBnlB
\left[
-\frac{77}{972}\right]
{+}\cfBtrAnlA
\left[
-\frac{169}{216} 
+\frac{11}{9}  \sbz \zeta_{3}
\right]
\nonumber\\
&{}& \hspace{1cm} {+}\cfAtrAnlAcaA
\left[
-\frac{769}{3888} 
-\frac{11}{9}  \sbz \zeta_{3}
\right]
{+}\cfBcaA
\left[
\frac{215}{216} 
-\frac{11}{18}  \sbz \zeta_{3}
\right]
\nnb
\\
&{}&\hspace{1cm}{+}\cfAcaB
\left[
\frac{5815}{15552} 
+\frac{11}{18}  \sbz \zeta_{3}
\right]
{},
\label{g3ns}
\end{eqnarray}

\begin{eqnarray}
{\g_4^{\sss NS} =  } 
&{}& \cfD
\left[
\frac{4157}{1536} 
+\frac{1}{2}  \sbz \zeta_{3}
\right]
{+}\dFFdRinvAnlA
\left[
-\frac{13}{12} 
-\frac{4}{3}  \sbz \zeta_{3}
+\frac{10}{3}  \sbz \zeta_{5}
\right]
\nonumber\\
&{+}&\dFARinvA
\left[
\frac{1}{4} 
-\frac{1}{3}  \sbz \zeta_{3}
-\frac{5}{3}  \sbz \zeta_{5}
\right]
{+}\cfAtrCnlC
\left[
\frac{107}{7776} 
+\frac{1}{54}  \sbz \zeta_{3}
\right]
\nonumber\\
&{+}&\cfBtrBnlB
\left[
\frac{4961}{10368} 
-\frac{119}{108}  \sbz \zeta_{3}
+\frac{11}{24}  \sbz \zeta_{4}
\right]
{+}\cfAtrBnlBcaA
\left[
-\frac{8191}{31104} 
+\frac{563}{432}  \sbz \zeta_{3}
-\frac{11}{24}  \sbz \zeta_{4}
\right]
\nonumber\\
&{+}&\cfCtrAnlA
\left[
\frac{2509}{1152} 
+\frac{67}{24}  \sbz \zeta_{3}
-\frac{145}{24}  \sbz \zeta_{5}
\right]
{+}\cfBtrAnlAcaA
\left[
-\frac{66451}{13824} 
+\frac{2263}{864}  \sbz \zeta_{3}
-\frac{143}{96}  \sbz \zeta_{4}
+\frac{85}{16}  \sbz \zeta_{5}
\right]
\nonumber\\
&{+}&\cfAtrAnlAcaB
\left[
\frac{22423}{31104} 
-\frac{9425}{1728}  \sbz \zeta_{3}
+\frac{143}{96}  \sbz \zeta_{4}
+\frac{15}{32}  \sbz \zeta_{5}
\right]
{+}\cfCcaA
\left[
-\frac{2585}{256} 
-\frac{71}{12}  \sbz \zeta_{3}
+\frac{935}{96}  \sbz \zeta_{5}
\right]
\nonumber\\
&{+}&\cfBcaB
\left[
\frac{882893}{82944} 
+\frac{11501}{3456}  \sbz \zeta_{3}
+\frac{121}{192}  \sbz \zeta_{4}
-\frac{715}{64}  \sbz \zeta_{5}
\right]
\nnb
\\
&{+}&\cfAcaC
\left[
-\frac{1192475}{497664} 
+\frac{5609}{3456}  \sbz \zeta_{3}
-\frac{121}{192}  \sbz \zeta_{4}
+\frac{275}{128}  \sbz \zeta_{5}
\right]
{},
\label{g4ns}
\end{eqnarray}

\bea
\g_3^{\sss SI} &=& \frac{d^{abc}\,d^{abc}}{d_R}  
\left\{
\frac{11}{144} 
-\frac{1}{6}  \sbz \zeta_{3}
\right\}
{},
\label{g3si}
\\
\g_4^{\sss SI} &=&  \frac{d^{abc}\,d^{abc}}{d_R} 
\Biggl\{
C_F
\left[
-\frac{13}{48} 
-\frac{1}{3}  \sbz \zeta_{3}
+\frac{5}{6}  \sbz \zeta_{5}
\right]
{+}\,C_A 
\left[
\frac{1015}{2304} 
-\frac{659}{768}  \sbz \zeta_{3}
+\frac{11}{64}  \sbz \zeta_{4}
+\frac{5}{64}  \sbz \zeta_{5}
\right]
\nonumber\\
&{}& \hspace{1cm} +\trAnlA
\left[
-\frac{55}{576} 
+\frac{41}{192}  \sbz \zeta_{3}
-\frac{1}{16}  \sbz \zeta_{4}
-\frac{5}{48}  \sbz \zeta_{5}
\right]
\Biggr\}
{}.
\label{g4si}
\end{eqnarray}

\section{QED $\beta$-functions in   five-loop order for different schemes}

\subsection{Massless QCD and QED: \MSbar-scheme}

The  polarization function $\Pi$ is known to be  directly related to the photon propagator, namely\footnote{
Without  loss of generality we will use the Landau gauge for the photon field.} 
\beq
{\cal  P }_{\al\be}(q) = \left( g_{\al\be} q^2 {-} q_{\al}\, q_{\be} \right)\,  \frac{d(Q^2)}{Q^{{4}}},
\ \ d(Q^2) = \frac{1}{1+ e^2\,\Pi^{\sss EM}(Q^2)}
{}.
\eeq
The combination   $e^2\, d(Q^2)$ is often referred to as ``invariant'' charge as it is renormalization 
scale {\em and scheme} independent due to the 
corresponding Ward identity. The independence of the invariant charge  on the renormalization scale $\mu$  directly leads to the 
RG equation  for the QED coupling constant $A(\mu) = \alpha(\mu)/(4\,\pi)=e(\mu)^2/(16\,\pi^2)$:
\beq
\mu^2 \frac{d}{d \mu^2} A = \beta^{\sss EM}(A,\as) = A^2\, (16 \pi^2)\, \g^{\sss EM}(a_s) = A^2 
\,d_R \, \sum_{i \ge 0}            \gamma_i^{\sss EM } \, \as^i
\label{QED:RG}
{},
\eeq
with 
\[
\g^{\sss EM}(a_s) \equiv \left(  \sum_i q_i^2 \right) \g^{\sss NS}
 + \left(  \sum_i q_i \right)^2 \g^{\sss SI} 
{},
\ \ 
\g^{\sss EM}_i \equiv  \left(  \sum_i q_i^2 \right) \g^{\sss NS}_i 
+ \left(  \sum_i q_i \right)^2 \g^{\sss SI}_i 
{}.
\]
The $\beta$-function $\beta^{\sss EM}(A,\as)$ describes the QCD-induced  corrections to the running of  $\alpha$ in
the \MSbar-scheme.

Using now  eqs.~(\ref{g0ns}-\ref{g4si}) and substituting the values 
of the colour factors corresponding  to the SU(3) colour group we find:
\bea
&{}& 
\beta^{\sss EM}(A,a_s) = A^2\left(  \sum_i q_i^2 \right)\left\{
4 +4\,a_s + 
a_s^2\,\Biggl( 
 \frac{125}{12}
-\frac{11}{18}\,n_f 
\Biggl)
\right. 
\nonumber\\
&{+}&
a_s^3
\Biggl(
\frac{10487}{432} 
+\frac{110}{9}  \sbz \zeta_{3}
{+} \,n_f 
\left[
-\frac{707}{216} 
-\frac{110}{27}  \sbz \zeta_{3}
\right]
\nonumber
-\frac{77}{972}\, n_f^2
\Biggr)
\nonumber\\
&{+}&
a_s^4
\Biggl(
\frac{2665349}{41472} 
+\frac{182335}{864}  \sbz \zeta_{3}
-\frac{605}{16}  \sbz \zeta_{4}
-\frac{31375}{288}  \sbz \zeta_{5}
\nonumber\\
&{+}& \,n_f 
\left[
-\frac{11785}{648} 
-\frac{58625}{864}  \sbz \zeta_{3}
+\frac{715}{48}  \sbz \zeta_{4}
+\frac{13325}{432}  \sbz \zeta_{5}
\right]
\nonumber\\
&{+}& \, n_f^2
\left[
-\frac{4729}{31104} 
+\frac{3163}{1296}  \sbz \zeta_{3}
-\frac{55}{72}  \sbz \zeta_{4}
\right]
{+} \, n_f^3
\left[
\frac{107}{15552} 
+\frac{1}{108}  \sbz \zeta_{3}
\right]
\Biggr)
\Biggr\}
+
\nonumber\\
&{+}& 
A^2\left(  \sum_i q_i \right)^2 \Biggl\{
a_s^3\Biggl(
\frac{55}{54} 
-\frac{20}{9}  \sbz \zeta_{3}
\Biggr)
{+}
a_s^4
\Biggl(
\frac{11065}{864} 
-\frac{34775}{864}  \sbz \zeta_{3}
+\frac{55}{8}  \sbz \zeta_{4}
+\frac{3875}{216}  \sbz \zeta_{5}
\nonumber\\
&{}&
\hspace{3cm}
+
\,n_f 
\left[
-\frac{275}{432} 
+\frac{205}{144}  \sbz \zeta_{3}
-\frac{5}{12}  \sbz \zeta_{4}
-\frac{25}{36}  \sbz \zeta_{5}
\right]
\Biggr)
\Biggr\}
\label{betaEM:MS}
{}.
\eea
For  the particular cases of 4, 5 and 6 quark flavours eq.~\re{betaEM:MS} takes the form
(the normalization is chosen in such a way to facilitate the comparison with \cite{Erler:1998sy})
\beq 
\beta^{\sss EM}(A,a_s)|_{n_f=4} = 4 \, A^2\left(
1.111 + 1.111 \, a_s + 2.214 \, a_s^2  + 1.210 \, a_s^3  - 3.904 \, a_s^4
\right)
{},
\eeq
\beq 
\beta^{\sss EM}(A,a_s)|_{n_f=5} = 4 \, A^2\left(
1.222 + 1.222 \, a_s + 2.249 \, a_s^2  - 1.227 \, a_s^3  - 13.429 \, a_s^4
\right)
{},
\eeq
\beq
\beta^{\sss EM}(A,a_s)|_{n_f=6} = 4 \, A^2\left(
1.667 + 1.667 \, a_s + 2.813 \, a_s^2- 5.791 \, a_s^3 - 32.336 a_s^4
\right)
{}.
\eeq

Another case of interest is pure QED, that is a theory  with $n_f$ single-charged fermions minimally coupled to the 
photon field. In this case the corresponding EM current is identical to the flavour singlet current. 
The corresponding $\beta$-function is, obviously,  obtained from the general  formula \re{QED:RG} by taking the QED values
for  the colour factors and setting $q_i=1$ and $a_s(\mu) = 4\,A (\mu)$:
\bea
&{}& \beta^{\sss QED}(A) = 
\,n_f 
\left[
\frac{4\ A^2}{3}  
\right]
{+}  
4\,  n_f A^3
-  A^4
\left[
2  \,n_f 
+\frac{44}{9}  \, n_f^2
\right]
\nonumber\\
&{+}&  A^5
\left[
-46  \,n_f 
+\frac{760}{27}  \, n_f^2
-\frac{832}{9}  \,\zeta_{3} \, n_f^2
-\frac{1232}{243}  \, n_f^3
\right]
\nonumber\\
&{+}& A^6\Biggl( \,n_f 
\left[
\frac{4157}{6} 
+128  \sbz \zeta_{3}
\right]
{+} \, n_f^2
\left[
-\frac{7462}{9} 
-992  \sbz \zeta_{3}
+2720  \sbz \zeta_{5}
\right]
\nonumber\\
&{+}& \, n_f^3
\left[
-\frac{21758}{81} 
+\frac{16000}{27}  \sbz \zeta_{3}
-\frac{416}{3}  \sbz \zeta_{4}
-\frac{1280}{3}  \sbz \zeta_{5}
\right]
{+} \, n_f^4
\left[
\frac{856}{243} 
+\frac{128}{27}  \sbz \zeta_{3}
\right]
\Biggr)
{}.
\eea 
If we set $n_f =1$, then the above result takes the form:
\bea
 \beta^{\sss QED}(A) &=& 
  \frac{4}{3} A^2
+  4 A^3
-\frac{62}{9} A^4
{+}  A^5
\left[
-\frac{5570}{243} 
-\frac{832}{9}  \sbz \zeta_{3}
\right]
\nonumber\\
&& \hspace{1cm} +\,  A^6
\left[
-\frac{195067}{486} 
-\frac{800}{3}  \sbz \zeta_{3}
-\frac{416}{3}  \sbz \zeta_{4}
+\frac{6880}{3}  \sbz \zeta_{5}
{}
\right]
\label{betaQED:nf:1}
\eea
or, numerically, 
\bea
\beta^{\sss QED}(A) &=&
 \frac{4}{3} A^2
+  4\,A^3
-\frac{62}{9} A^4
{+}  A^5\,
(-7.116 - 126.93^{\sss SI}) + 
 A^6\,(776.39 + 729.63^{\sss SI})
\nnb
\\
&=&
 \frac{4}{3} A^2
+ 4\, A^3
-\frac{62}{9} A^4
- 134.045 \, A^5 +  1506.02\,A^6
{},
\eea
where in the first line  we have explicitly separated  non-singlet from singlet contributions.

For future reference it is also useful to present the evolution equation for the QED polarization function:
\beq
{
\frac{\prd }{\prd L}\,  \Pi^{\sss QED} (L,A) =
\g^{\sss QED}(A)
-
 \beta^{\sss QED}(A) \frac{\prd }{\prd A}
\,
\Pi^{\sss QED}(L,A)
\label{rgPi2:QED}
{}.
}
\eeq
Here
\bea 
\beta^{\sss QED}(A) & \equiv &  A^2\, (16 \pi^2)\, \g^{\sss QED}(A)
{},
\label{betaQED:def}
\\
\Pi^{\sss QED}(L,A) &\equiv&  n_f\,\Pi^{\sss NS,QED}(L,A) + n_f^2\, \Pi^{\sss SI,QED}(L,A)
{}.
\label{PiQED:def}
\eea
In addition,  $\Pi^{\sss NS,QED}$, $\Pi^{\sss SI,QED}$ and $\g^{\sss QED}$ are $\Pi^{\sss NS}$, $\Pi^{\sss SI}$ and $\g$ respectively 
with  all  colour factors substituted  according to eq.~\re{U1} and $a_s=4\,A$.

\subsection{Massless QCD and QED: MOM-scheme}

The MOM-scheme for   the QED coupling constant is defined by the requirement that
at $Q^2 = \mu^2 $ the polarization function would vanish.  The scheme  independence of the invariant charge
directly leads to the following  relation between \MSbar-renormalized QED coupling constant $A(\mu)$ and
its MOM analog 
\newcommand{\AMOM}{\tilde{A}}
\beq
 \AMOM (\mu) = \frac{ A(\mu)}{1+ (4\pi)^2\,A(\mu) \, \Pi(L=0,a_s(\mu))} 
\label{AMOM:QCD}
{}.
\eeq
The above equation allows to express easily the QED $\beta$-function in
the MOM-scheme via the corresponding \MSbar \ $\beta$-function and the
perturbative expansion of the polarization function as given in 
eqs.~(\ref{PiEM},\ref{p0ns}--\ref{p3si},\ref{betaEM:MS}). 
After differentiating the rhs of \re{AMOM:QCD} with respect to $\mu$ and
using  eq.~\re{rgea}  we obtain 
\bea
\beta^{\sss EM}(\AMOM,a_s) &=&  16\pi^2\,\AMOM^2\,\Biggl[
\g^{\sss EM}(a_s) - \beta(a_s)\,  \frac{\prd}{\prd a_s}\, \Pi^{\sss EM}(L=0,a_s)
\Biggr]
\nnb
\\
&=& \frac{4}{3}\AMOM^2\,D^{\sss EM}(L=0,a_s)
\label{betaMOMqcd}
{},
\eea
where 
\beq
{D}^{\sss EM}(L,a_s) =  -12\, \pi^2
Q^2\, \frac{\mathrm{d}}{\mathrm{d} Q^2}\Pi^{\sss EM}(L=\ln\frac{\mu^2}{Q^2},a_s)
{}.
\label{D:EM}
\eeq
Explicitly, for the $SU(3)$ colour group  we get
\bea
&{}& 
\beta^{\sss EM}(\AMOM,a_s) = \frac{4\,\AMOM^2}{3}\,\left(  3\,\sum_i q_i^2 \right)\Biggl\{
 1+ a_s 
+
a_s^2\,\Biggl(
\frac{365}{24} 
-11  \sbz \zeta_{3}
+ \,n_f 
\left[
-\frac{11}{12} 
+\frac{2}{3}  \sbz \zeta_{3}
\right]
\Biggr)
\nonumber\\
&{+}& 
a_s^3\,
\Biggl(
\frac{87029}{288} 
-\frac{1103}{4}  \sbz \zeta_{3}
+\frac{275}{6}  \sbz \zeta_{5}
+ \,n_f 
\left[
-\frac{7847}{216} 
+\frac{262}{9}  \sbz \zeta_{3}
-\frac{25}{9}  \sbz \zeta_{5}
\right]
{+} \, n_f^2
\left[
\frac{151}{162} 
-\frac{19}{27}  \sbz \zeta_{3}
\right]
\Biggr)
\nonumber\\
&{+}& 
a_s^4\,
\Biggl(
\frac{144939499}{20736} 
-\frac{5693495}{864}  \sbz \zeta_{3}
+\frac{5445}{8}  \,\zeta_3^2
+\frac{65945}{288}  \sbz \zeta_{5}
-\frac{7315}{48}  \sbz \zeta_{7}
\nonumber\\
&{+}& 
\,n_f 
\left[
-\frac{13044007}{10368} 
+\frac{12205}{12}  \sbz \zeta_{3}
-55  \,\zeta_3^2
+\frac{29675}{432}  \sbz \zeta_{5}
+\frac{665}{72}  \sbz \zeta_{7}
\right]
\nonumber\\
&{+}& \, n_f^2
\left[
\frac{1045381}{15552} 
-\frac{40655}{864}  \sbz \zeta_{3}
+\frac{5}{6}  \,\zeta_3^2
-\frac{260}{27}  \sbz \zeta_{5}
\right]
+ \, n_f^3
\left[
-\frac{6131}{5832} 
+\frac{203}{324}  \sbz \zeta_{3}
+\frac{5}{18}  \sbz \zeta_{5}
\right]
\Biggr)\Biggr\}
\nonumber\\
&{+}& 
\frac{4\,\AMOM^2}{3}\left(  \sum_i q_i \right)^2 \Biggl\{
a_s^3\Biggl(
\frac{55}{72} 
-\frac{5}{3}  \sbz \zeta_{3}
\Biggl)
+
a_s^4\Biggl(
\frac{5795}{192} 
-\frac{8245}{144}  \sbz \zeta_{3}
-\frac{55}{4}  \,\zeta_3^2
+\frac{2825}{72}  \sbz \zeta_{5}
\nonumber\\
&{+}& \,n_f 
\left[
-\frac{745}{432} 
+\frac{65}{24}  \sbz \zeta_{3}
+\frac{5}{6}  \,\zeta_3^2
-\frac{25}{12}  \sbz \zeta_{5}
\right]
\biggr)
\Biggr\}
\label{betaEM:MOM}
{}.
\eea
For  particular cases of 4, 5 and 6 quark flavours eq.~\re{betaEM:MOM}  reads
\beq 
\beta^{\sss EM}(\AMOM,a_s)|_{n_f=4} = 4 \, \AMOM^2\left(
1.111 + 1.111 \,a_s + 1.694 \,a_s^2  + 2.881 \,a_s^3  + 28.132 \,a_s^4
\right)
{},
\eeq
\beq 
\beta^{\sss EM}(\AMOM,a_s)|_{n_f=5} = 4 \, \AMOM^2\left(
1.222 + 1.222 \,a_s + 1.723 \,a_s^2   - 0.879 \,a_s^3  + 10.703 \,a_s^4
\right)
{},
\eeq
\beq
\beta^{\sss EM}(\AMOM,a_s)|_{n_f=6} = 4 \, \AMOM^2\left(
1.667 + 1.667 \,a_s + 2.156 \,a_s^2   - 6.995 \,a_s^3  - 13.994 \,a_s^4
\right)
{}.
\eeq

In the same way one could derive the pure QED $\beta$-function $\beta^{\sss QED}(\AMOM)$ 
in the MOM-scheme (it was first introduced by Gell-Mann and Low in \cite{GellMann:1954fq}
under the name  ``$\psi$ function'').
Indeed,  an analog of eq.~\re{AMOM:QCD} now assumes  the following form:
\beq
 \AMOM(\mu) = \frac{ A(\mu)}{1+ (4\pi)^2\,A(\mu) \, \Pi^{\sss QED}(L=0,A)} 
\label{AMOM:QED}
{}.
\eeq
The resulting formula  for the QED $\beta$-function  in the MOM-scheme reads
(we have  used relation \re{rgPi2:QED} and definitions (\ref{betaQED:def},\ref{PiQED:def}))
\bea
\beta^{\sss QED}(\AMOM) &=&  16\pi^2\,\AMOM^2\,\Biggl[
\g^{\sss QED}(A) - \beta^{\sss QED}(A)\,  \frac{\prd}{\prd A}\, \Pi^{\sss QED}(L=0,A)
\Biggr]
\nnb
\\
&=& \frac{4}{3}\AMOM^2\,D^{\sss QED}(L=0,A)
\label{betaMOMqed}
{},
\eea
where 
$D^{\sss QED}(L,A) \equiv -12\pi^2 Q^2\, \frac{\mathrm{d}}{\mathrm{d} Q^2} \,\Pi^{\sss QED}(L=\ln\frac{\mu^2}{Q^2},A)
{}.
$ 
In addition, the coupling constant $A(\mu)$ appearing in the rhs of eq.~\re{betaMOMqed}
should be  expressed in terms of $\AMOM(\mu)$ using an inversion  of eq.~\re{AMOM:QED}.
Explicitly,  for the U(1) gauge group we arrive at the following result:
\bea
&{}& \beta^{\sss QED}(\AMOM) = 
\,n_f 
\left[
\frac{4\ \AMOM^2}{3}  
\right]
{+}  
4\,  n_f \AMOM^3
+  \AMOM^4
\Biggl(
-2  \,n_f 
+ n_f^2\left[
-\frac{184}{9}  + \frac{64}{3}\zeta_3
\right]
\Biggr)
\nonumber\\
&{+}&  \AMOM^5
\Biggl(
-46 \,n_f
+ n_f^2
\left[
104 
+\frac{512}{3}  \sbz \zeta_{3}
-\frac{1280}{3}  \sbz \zeta_{5}
\right]
{+} \, n_f^3
\left[
128 
-\frac{256}{3}  \sbz \zeta_{3}
\right]
\Biggr)
\nonumber\\
&{+}& \AMOM^6 \Biggl(
 \,n_f 
\left[
\frac{4157}{6} 
+128  \sbz \zeta_{3}
\right]
{+} \, n_f^2
\left[
-1004 
-\frac{2944}{3}  \sbz \zeta_{3}
-5760  \sbz \zeta_{5}
+8960  \sbz \zeta_{7}
\right]
\label{betaQED:MOM}
\\
&{+}& \, n_f^3
\left[
-\frac{27064}{27} 
-\frac{26240}{9}  \sbz \zeta_{3}
+1024  \,\zeta_3^2
+2560  \sbz \zeta_{5}
\right]
{+} \, n_f^4
\left[
-\frac{8576}{9} 
+\frac{3584}{9}  \sbz \zeta_{3}
+\frac{5120}{9}  \sbz \zeta_{5}
\right]
\Biggr)
\nnb
{}.
\eea 
After setting  $n_f=1$ we get:
\bea
 \beta^{\sss QED}(\AMOM) &=& 
  \frac{4}{3} \AMOM^2
+ 4\,  \AMOM^3
+
 \AMOM^4
\left[
-\frac{202}{9} 
+\frac{64}{3}  \sbz \zeta_{3}
\right]
{+}  \AMOM^5
\left[
186 
+\frac{256}{3}  \sbz \zeta_{3}
-\frac{1280}{3}  \sbz \zeta_{5}
\right]
\nonumber\\
&{+}&  \AMOM^6
\left[
-\frac{122387}{54} 
-\frac{10112}{3}  \sbz \zeta_{3}
+1024  \,\zeta_3^2
-\frac{23680}{9}  \sbz \zeta_{5}
+8960  \sbz \zeta_{7}
\right]
\nnb
\\
&=&
 \frac{4}{3} A^2
+  4\,A^3
+ 3.199 \, A^4
{+}  A^5\,
(-26.918 - 126.929^{\sss SI}) + 
 A^6\,(1054.41 + 413.592^{\sss SI})
\nnb
\\
&=&
 \frac{4}{3} A^2
+  4\,A^3
+ 3.199\, A^4 
 - 153.847 A^5\,
+ 1467.998\, \,A^6
{},
\label{betaQED:MOM:N}
\eea
where the singlet contribution  has again been  identified separately.

{

\section{Discussion}

\begin{table}[h]
\begin{center}
\begin{tabular}{|c|c|c|c|c|}
\hline 
 $\ell$  & 1  & 2 & 3 & 4  \\
\hline
&- & $\zeta_3$ & $\zeta_3, \zeta_4, \zeta_5$ &  $\zeta_3, \zeta_4, \zeta_5, \zeta_3^2, \zeta_6, \zeta_7$
\\
\hline
\end{tabular}
\caption{
\label{table1}
Possible irrational structures which are allowed to appear in $\ell$-loop massless propagators.
}
\end{center}
\end{table}
Let us  discuss the structure of 
trancendentalities appearing in our results.
It follows from work \cite{Baikov:2010hf} that  the variety of  $\zeta$-constants entering into
the $\ovl{\mbox{MS}}$-renormalized (euclidian) massless propagators
 depends  on the loop order according to Table \ref{table1}.
Table \ref{table2} provides  the same information about possible irrational numbers which could show up in 
anomalous dimensions. 
\begin{table}[h]
\begin{center}
\begin{tabular}{|c|c|c|c|c|}
\hline 
 $\ell$  & 1,2  & 3 & 4 & 5  \\
\hline
&- & $\zeta_3$ & $\zeta_3, \zeta_4, \zeta_5$ &  $\zeta_3, \zeta_4, \zeta_5, \zeta_3^2, \zeta_6, \zeta_7$
\\
\hline
\end{tabular}
\caption{
\label{table2}
Possible irrational structures which are allowed to appear in $\ell$-loop anomalous dimensions and $\beta$-functions.
}
\end{center}
\end{table}
An examination of  eqs.~(\ref{p0ns}--\ref{p3si}, \ref{g0ns}--\ref{g4si}) immediately reveals 
that the observed  pattern of trancendentalities is significantly more limited 
than what is allowed by Tables \ref{table1} and \ref{table2}. Indeed,  the four-loop anomalous dimension $\g_3$ 
contains no $\zeta_4$ and no $\zeta_5$  while the three-loop polarization function contains  $\zeta_5$ but does not 
comprise  $\zeta_4$. 
Let us  move up one loop. The situation is getting even more puzzling:  the five-loop anomalous dimension $\g_4$ 
does contain $\zeta_4$ but still does not include  $\zeta_3^2, \zeta_6$ and $ \zeta_7$. The 
four-loop polarization function contains $\zeta_4$ but is  free from $\zeta_6$. 
Unfortunately, we are not aware about the existence of  any reason  behind these  remarkable facts, except for 
 one observation, namely the absence of $\zeta_4$ in the MOM $\beta$-functions \re{betaEM:MOM}} and \re{betaQED:MOM}. 

Indeed, according to eqs.~\re{betaMOMqcd} and \re{betaMOMqed} 
the constant $\zeta_4$ does not appear in these two $\beta$-functions since it  
does not appear in the Adler function. 
However, the puzzle  of the  absence of $\zeta_4$ in ${\cal O}(\alpha_s^3)$ contribution to the Adler function 
has been recently fully  explained  in  \cite{Baikov:2010hf}. The explanation is based on a quite peculiar structure
of irrational contributions to each  four-loop master integral.  

Why  this absence continues to  hold at five loops  is still  a mystery (at least for us). In all probability it 
is connected  with some regularities of {\em five-loop} master integrals. 
But here starts  terra incognita $\dots$

\section{Conclusions}

We have presented four  new results, namely the QED  $\beta$-functions
$\beta^{\sss EM}$ and $\beta^{\sss QED}$ in the   \MSbar- \   and MOM-schemes.

We have described the  status of results for the vector correlator in
massless QCD. These  are not completely new as they
have been used to produce the Adler function and $R(s)$ in works
 \cite{Baikov:2008jh,Baikov:2010je,Baikov:2012er,Baikov:dummy}. Nevertheless, we believe that the separate presentation of
the polarization function and its anomalous dimension is both  useful and
instructive.  
First,  it reflects   the real way how the calculations have been done.
Second, it  clearly demonstrates puzzling regularities of the structure of
irrational terms contributing to $\Pi$ and $\gamma$. 
Third,  it  makes trivial the  construction   of the QED $\beta$-function in the  \MSbar-  \ and MOM-schemes
(cmp. with the somewhat unnecessary complicated ``inverse engineering''  employed in \cite{Erler:1998sy}
to reconstruct the function $\beta^{\sss EM}(A,a_s)$ at  four loop level).

The calculations of $\Pi$ and $\g$  have  been performed
on a SGI ALTIX 24-node IB-interconnected cluster of 8-cores Xeon
computers 
using  parallel  MPI-based \cite{Tentyukov:2004hz} as well as thread-based 
\cite{Tentyukov:2007mu} versions  of FORM
\cite{Vermaseren:2000nd}.  For the evaluation of colour factors we have used the FORM program { COLOR}
\cite{vanRitbergen:1998pn}. The diagrams have been generated with QGRAF \cite{Nogueira:1991ex}.
The figures have been drawn with the   help of
Axodraw \cite{Vermaseren:1994je} and JaxoDraw  \cite{Binosi:2003yf}.

This work was supported by the Deutsche Forschungsgemeinschaft in the
Sonderforschungsbereich/Transregio SFB/TR-9 ``Computational Particle
Physics'' and  by RFBR grants  11-02-01196, 10-02-00525.

Finally, we  want to note that the result \re{betaQED:nf:1} for the   
function $\beta^{\sss QED}(A)$  with \mbox{$n_f =1$}   was first
reported  by one of the present authors in  September 2011 during the
10-th International Symposium on Radiative Corrections, 
RADCOR 2011 (see the 10-th  page of the file 
{\tt http://www.icts.res.in/media/uploads/Program/Files/chet.pdf}).

\providecommand{\href}[2]{#2}\begingroup\raggedright\endgroup

\ed

\bibliographystyle{JHEP}

\bibliography{dim_reg,JJ,chet09,chet,steinhauser,baikov,asmirnov,smirnov,vladimirov,vermaseren,%
surguladze,laporta,gorishnii,tarasov,bierenbaum,kazakov,david_dirk_qQED,other_masters,%
sector_decom,lee,kotikov_before1995,remiddi_1997-2000,broadhurst,%
DIS,DIS2,DIS4}

\begin{thebibliography}{10}

\bibitem{Aoyama:2012wj}
T.~Aoyama, M.~Hayakawa, T.~Kinoshita, and M.~Nio, {\it {Tenth-Order QED
  Contribution to the Electron g-2 and an Improved Value of the Fine Structure
  Constant}},  \href{http://xxx.lanl.gov/abs/1205.5368}{{\tt arXiv:1205.5368}}.

\bibitem{Karshenboim:2005iy}
S.~G. Karshenboim, {\it {Precision physics of simple atoms: QED tests, nuclear
  structure and fundamental constants}},  {\em Phys. Rept.} {\bf 422} (2005)
  1--63, [\href{http://xxx.lanl.gov/abs/hep-ph/0509010}{{\tt hep-ph/0509010}}].

\bibitem{Eides:2007}
M.~I. Eides, H.~Grotch, and V.~A. Shelyuto, {\it Theory of light hydrogenic
  bound states},  {\em Springer Tracts Mod. Phys.} {\bf 222} (2007) 1--262.

\bibitem{Stueckelberg53}
E.~Stueckelberg and A.~Petermann, {\it {La normalisation des constantes dans la
  theorie des quanta}},  {\em Helv. Phys. Acta.} {\bf 26} (1953) 499--520.

\bibitem{GellMann:1954fq}
M.~Gell-Mann and F.~Low, {\it {Quantum electrodynamics at small distances}},
  {\em Phys.Rev.} {\bf 95} (1954) 1300--1312.

\bibitem{Bogolyubov:1956gh}
N.~Bogolyubov and D.~Shirkov, {\it {Charge renormalization group in quantum
  field theory}},  {\em Nuovo Cim.} {\bf 3} (1956) 845--863.

\bibitem{Adler:1974gd}
S.~L. Adler, {\it Some simple vacuum polarization phenomenology: $e^+ e^- \to $
  hadrons: The mu - mesic atom x-ray discrepancy and (g-2) of the muon},  {\em
  Phys. Rev.} {\bf D10} (1974) 3714.

\bibitem{Chetyrkin:1979bj}
K.~G. Chetyrkin, A.~L. Kataev, and F.~V. Tkachov, {\it {Higher Order
  Corrections to {$\sigma_{tot}(e^+ e^- \to \mbox{Hadrons})$} in Quantum
  Chromodynamics}},  {\em Phys. Lett.} {\bf B85} (1979) 277.

\bibitem{Gorishnii:1990vf}
S.~G. Gorishny, A.~L. Kataev, and S.~A. Larin, {\it {The O (alpha-s**3)
  corrections to sigma-tot (e+ e- $\to$ hadrons) and Gamma (tau- $\to$
  tau-neutrino + hadrons) in QCD}},  {\em Phys. Lett.} {\bf B259} (1991)
  144--150.

\bibitem{Surguladze:1990tg}
L.~R. Surguladze and M.~A. Samuel, {\it {Total hadronic cross-section in e+ e-
  annihilation at the four loop level of perturbative QCD}},  {\em Phys. Rev.
  Lett.} {\bf 66} (1991) 560--563.

\bibitem{Baikov:2001aa}
P.~Baikov, K.~Chetyrkin, and J.~H. Kuhn, {\it {The Cross section of e+ e-
  annihilation into hadrons of order alpha(s)**4 n(f)**2 in perturbative QCD}},
   {\em Phys.Rev.Lett.} {\bf 88} (2002) 012001,
  [\href{http://xxx.lanl.gov/abs/hep-ph/0108197}{{\tt hep-ph/0108197}}].

\bibitem{Baikov:2008jh}
P.~A. Baikov, K.~G. Chetyrkin, and J.~H. K{\"u}hn, {\it {Order $\alpha^4_s$ QCD
  Corrections to $Z$ and $\tau$ Decays}},  {\em Phys. Rev. Lett.} {\bf 101}
  (2008) 012002, [\href{http://xxx.lanl.gov/abs/0801.1821}{{\tt
  arXiv:0801.1821}}].

\bibitem{Baikov:2010je}
P.~A. Baikov, K.~G. Chetyrkin, and J.~H. K{\"u}hn, {\it {Adler Function,
  Bjorken Sum Rule, and the Crewther Relation to Order $alpha_s^4$ in a General
  Gauge Theory}},  {\em Phys. Rev. Lett.} {\bf 104} (2010) 132004,
  [\href{http://xxx.lanl.gov/abs/1001.3606}{{\tt arXiv:1001.3606}}].

\bibitem{Baikov:2012er}
P.~A. Baikov, K.~G. Chetyrkin, J.~H. K\"uhn, and J.~Rittinger, {\it {Complete
  ${\cal O}(\alpha_{s}^{4})$ QCD Corrections to Hadronic $Z$ Decays}},  {\em
  Phys. Rev. Lett.} {\bf 108} (May, 2012) 222003.

\bibitem{Baikov:dummy}
P.~A. Baikov, K.~G. Chetyrkin, J.~H. K{\"u}hn, and J.~Rittinger, {\it {Adler
  Function, Sum Rules and Crewther Relation of Order ${\cal O}(\alpha_s^4)$:
  the Singlet Case}},  {\em in preparation} (2012) {},
  [\href{http://xxx.lanl.gov/abs/hep-ph/12}{{\tt hep-ph/12}}].

\bibitem{Baikov:2008cp}
P.~A. Baikov, K.~G. Chetyrkin, and J.~H. K{\"u}hn, {\it {Massless propagators:
  applications in QCD and QED}},  {\em PoS} {\bf RADCOR2007} (2007) 023,
  [\href{http://xxx.lanl.gov/abs/0810.4048}{{\tt arXiv:0810.4048}}].

\bibitem{Chetyrkin:1996ia}
K.~G. Chetyrkin, J.~H. K{\"u}hn, and A.~Kwiatkowski, {\it {QCD corrections to
  the $e^{+} e^{-}$ cross-section and the $Z$ boson decay rate: Concepts and
  results}},  {\em Phys. Rept.} {\bf 277} (1996) 189--281.

\bibitem{Gorishnii:1990kd}
S.~G. Gorishny, A.~L. Kataev, S.~A. Larin, and L.~R. Surguladze, {\it {The
  Analytical four loop corrections to the QED Beta function in the MS scheme
  and to the QED psi function: Total reevaluation}},  {\em Phys. Lett.} {\bf
  B256} (1991) 81--86.

\bibitem{Erler:1998sy}
J.~Erler, {\it {Calculation of the QED coupling alpha (M(Z)) in the modified
  minimal subtraction scheme}},  {\em Phys.Rev.} {\bf D59} (1999) 054008,
  [\href{http://xxx.lanl.gov/abs/hep-ph/9803453}{{\tt hep-ph/9803453}}].

\bibitem{Tarasov:1980au}
O.~V. Tarasov, A.~A. Vladimirov, and A.~Y. Zharkov, {\it The gell-mann-low
  function of qcd in the three loop approximation},  {\em Phys. Lett.} {\bf
  B93} (1980) 429--432.

\bibitem{Larin:1993tp}
S.~A. Larin and J.~A.~M. Vermaseren, {\it {The Three loop QCD Beta function and
  anomalous dimensions}},  {\em Phys. Lett.} {\bf B303} (1993) 334--336,
  [\href{http://xxx.lanl.gov/abs/hep-ph/9302208}{{\tt hep-ph/9302208}}].

\bibitem{Chetyrkin:1984xa}
K.~G. Chetyrkin and V.~A. Smirnov, {\it {R* OPERATION CORRECTED}},  {\em Phys.
  Lett.} {\bf B144} (1984) 419--424.

\bibitem{Chetyrkin:1996ez}
K.~G. Chetyrkin, {\it {Corrections of order alpha(s)**3 to R(had) in pQCD with
  light gluinos}},  {\em Phys. Lett.} {\bf B391} (1997) 402--412,
  [\href{http://xxx.lanl.gov/abs/hep-ph/9608480}{{\tt hep-ph/9608480}}].

\bibitem{Baikov:2005nv}
P.~A. Baikov, {\it A practical criterion of irreducibility of multi-loop
  feynman integrals},  {\em Phys. Lett.} {\bf B634} (2006) 325--329,
  [\href{http://xxx.lanl.gov/abs/hep-ph/0507053}{{\tt hep-ph/0507053}}].

\bibitem{Baikov:1996rk}
P.~A. Baikov, {\it {Explicit solutions of the 3--loop vacuum integral
  recurrence relations}},  {\em Phys. Lett.} {\bf B385} (1996) 404--410,
  [\href{http://xxx.lanl.gov/abs/hep-ph/9603267}{{\tt hep-ph/9603267}}].

\bibitem{Baikov:2010hf}
P.~A. Baikov and K.~G. Chetyrkin, {\it {Four Loop Massless Propagators: an
  Algebraic Evaluation of All Master Integrals}},  {\em Nucl. Phys.} {\bf B837}
  (2010) 186--220, [\href{http://xxx.lanl.gov/abs/1004.1153}{{\tt
  arXiv:1004.1153}}].

\bibitem{Lee:2011jt}
R.~N. Lee, A.~V. Smirnov, and V.~A. Smirnov, {\it {Master Integrals for
  Four-Loop Massless Propagators up to Transcendentality Weight Twelve}},  {\em
  Nucl. Phys.} {\bf B856} (2012) 95--110,
  [\href{http://xxx.lanl.gov/abs/1108.0732}{{\tt arXiv:1108.0732}}].

\bibitem{Vermaseren:1997fq}
J.~A.~M. Vermaseren, S.~A. Larin, and T.~van Ritbergen, {\it The 4-loop quark
  mass anomalous dimension and the invariant quark mass},  {\em Phys. Lett.}
  {\bf B405} (1997) 327--333,
  [\href{http://xxx.lanl.gov/abs/hep-ph/9703284}{{\tt hep-ph/9703284}}].

\bibitem{Tentyukov:2004hz}
M.~Tentyukov {\em et.~al.}, {\it {ParFORM: Parallel Version of the Symbolic
  Manipulation Program FORM}},  \href{http://xxx.lanl.gov/abs/cs/0407066}{{\tt
  cs/0407066}}.

\bibitem{Tentyukov:2007mu}
M.~Tentyukov and J.~A.~M. Vermaseren, {\it {The multithreaded version of
  FORM}},  \href{http://xxx.lanl.gov/abs/hep-ph/0702279}{{\tt hep-ph/0702279}}.

\bibitem{Vermaseren:2000nd}
J.~A.~M. Vermaseren, {\it New features of form},
  \href{http://xxx.lanl.gov/abs/math-ph/0010025}{{\tt math-ph/0010025}}.

\bibitem{vanRitbergen:1998pn}
T.~van Ritbergen, A.~N. Schellekens, and J.~A.~M. Vermaseren, {\it Group theory
  factors for feynman diagrams},  {\em Int. J. Mod. Phys.} {\bf A14} (1999)
  41--96, [\href{http://xxx.lanl.gov/abs/hep-ph/9802376}{{\tt
  hep-ph/9802376}}].

\bibitem{Nogueira:1991ex}
P.~Nogueira, {\it Automatic feynman graph generation},  {\em J. Comput. Phys.}
  {\bf 105} (1993) 279--289.

\bibitem{Vermaseren:1994je}
J.~A.~M. Vermaseren, {\it Axodraw},  {\em Comput. Phys. Commun.} {\bf 83}
  (1994) 45--58.

\bibitem{Binosi:2003yf}
D.~Binosi and L.~Theussl, {\it {JaxoDraw: A graphical user interface for
  drawing Feynman diagrams}},  {\em Comput. Phys. Commun.} {\bf 161} (2004)
  76--86, [\href{http://xxx.lanl.gov/abs/hep-ph/0309015}{{\tt
  hep-ph/0309015}}].

\end{thebibliography}

\ed